\documentclass[fleqn,10pt]{wlscirep}
\usepackage[utf8]{inputenc}
\usepackage[T1]{fontenc}
\usepackage{multirow, makecell}
\usepackage{bm}
\usepackage{siunitx}

\newcommand{\black}{\color{black}}

\title{Numerical study of electron acceleration in structured CNT targets via self-injection in a wakefield bubble driven by an 800 nm laser}

\author[1,*]{Cristian Bon\c toiu}
\author[2,*]{Alexandre Bonatto}
\author[3,4]{{\"O}znur Apsimon}
\author[5]{Laura Bandiera}
\author[6,7]{Gianluca Cavoto}
\author[8]{Illya Drebot}
\author[9]{Giancarlo Gatti}
\author[10]{Jorge Giner-Navarro}
\author[1,4]{Bifeng Lei}
\author[11]{Pablo Martín-Luna}
\author[6,7]{Ilaria Rago}
\author[10]{Juan Rodríguez Pérez}
\author[12]{Bruno Silveira Nunes}
\author[5]{Alexei Sytov}
\author[13]{Constantinos Valagiannopoulos}
\author[1,4]{Carsten P. Welsch}
\author[3,4]{Guoxing Xia}
\author[3,4]{Jiaqi Zhang}
\author[10,14,*]{Javier Resta-López}

\affil[1]{Department of Physics, University of Liverpool, Oxford Street, Liverpool L69~7ZE, UK}

\affil[2]{Graduate Program in Information Technology and Healthcare Management, and the Beam Physics Group, Federal University of Health Sciences of Porto Alegre, Porto Alegre, RS, 90050-170, Brazil}

\affil[3]{Department of Physics and Astronomy, University of Manchester, Oxford Road, Manchester, M13 9PL, UK}

\affil[4]{Cockroft Institute, Keckwick Lane, Daresbury, Warrington WA4 4AD, UK}

\affil[5]{INFN Sezione di Ferrara, Via Saragat 1, 44122 Ferrara, Italy}

\affil[6]{Dipartimento di Fisica, Sapienza Università di Roma, Piazzale Aldo Moro 2, 00185 Roma, Italy}

\affil[7]{INFN Sezione di Roma, Piazzale Aldo Moro 2, 00185 Rome, Italy}

\affil[8]{INFN Sezione di Milano, Via G. Celoria 16, 20133 Milano, Italy}

\affil[9]{Centro de Láseres Pulsados, Building M5, Science Park, Calle Adaja 8, 37185 Villamayor, Salamanca, Spain}

\affil[10]{Instituto de Ciencia de los Materiales (ICMUV), Universidad
de Valencia, 46071 Paterna, Spain}

\affil[11]{Instituto de Física Corpuscular (IFIC), Universidad de Valencia - CSIC, 46980 Paterna, Spain}

\affil[12]{Nuclear and Energy Research Institute, IPEN–CNEN, São Paulo, SP, CEP 05508-000, Brazil}

\affil[13]{National Technical University of Athens, School of Electrical and Computer Engineering,  Zografou 157 73, Greece}

\affil[14]{Departamento de F\'isica Aplicada y Electromagnetismo, Universidad de Valencia, 46100 Burjassot, Spain}

\affil[*]{cbontoiu@gmail.com, abonatto@ufcspa.edu.br, javier2.resta@uv.es}

\begin{abstract}

Laser wakefield acceleration (LWFA) may achieve TeV/m gradients using high-density solid-state plasmas as accelerating media. However, the application of bulk solid materials requires attosecond laser pulses, such as X-ray lasers, to drive wakefields at these high densities. Additionally, the short wakefield wavelengths associated with solid-state plasmas greatly limit the accelerating length. An alternative approach employs 2D carbon-based nanomaterials, like graphene or carbon nanotubes (CNTs), configured into structured targets. These nanostructures are designed with voids or low-density regions to effectively reduce the overall plasma density. This reduction enables the use of longer-wavelength lasers and also extends the plasma wavelength and the acceleration length. 
In this study, we present, to our knowledge, the first numerical demonstration of electron acceleration via self-injection into a wakefield bubble driven by an infrared laser pulse in structured CNT targets, similar to the behavior observed in gaseous plasmas for LWFA in the nonlinear (or bubble) regime. Using the PIConGPU code, bundles of CNTs are modeled in a 3D geometry as 25 nm-thick carbon tubes with an initial density of $10^{22} \, \text{cm}^{-3}$. The carbon plasma is ionized by a three-cycle, 800 nm wavelength laser pulse with a peak intensity of $10^{21} \, \text{W cm}^{-2}$, achieving an effective plasma density of $10^{20} \, \text{cm}^{-3}$. The same laser also drives the wakefield bubble, responsible for the electron self-injection and acceleration. Simulation results indicate that fs-long electron bunches with hundreds of pC charge can be self-injected and accelerated at gradients exceeding 1 TeV/m. Both charge and accelerating gradient figures are unprecedented when compared with LWFA in gaseous plasma.
\end{abstract}

\begin{document}
\flushbottom
\maketitle 
\section*{Introduction}

Over the past few decades, the field of plasma-based accelerators has made remarkable advances. The long-ago predicted GeV / m accelerating gradients~\cite{Tajima1979,Chen1985,esarey_2009} are now routinely achieved by multiple experimental facilities around the globe~\cite{Gonsalves2019,DArcy2019}, paving the way for the design of future compact accelerators with applications that range from medical~\cite{Labate2020} and industrial~\cite{Assmann2020} uses to high-energy physics research~\cite{Schroeder2010,Schroeder2016}. Moreover, because the accelerating gradients are proportional to the plasma density, researchers are exploring the use of solid-state plasmas to push these gradients even higher. Solid-state plasmas have densities on the order of $10^{22}$ to $10^{24}\,\text{cm}^{-3}$, which could potentially enable $\si{TeV/m}$ gradients. Attaining such high gradients would be a significant leap forward in accelerator technology. However, achieving plasma-based acceleration at these high densities presents significant challenges. The extremely short plasma wavelengths associated with such densities require sub-micrometer quasi-solid electron beams~\cite{Yakimenko2019} or attosecond X-ray laser pulses~\cite{Tajima2014} to be adopted as drivers for the accelerating wakefields. Drivers with these characteristics are either not yet available or remain extremely limited in availability. An alternative approach to overcome these challenges involves the use of nanostructured targets. By arranging bundles of nanomaterials—such as carbon nanotubes (CNTs) or graphene layers—in an alternating pattern with empty or low-density regions, the effective density of the target can be adjusted. As the driver propagates through these empty regions, it ionizes the adjacent solid walls, with the ionized electrons populating the gaps and forming a high-density plasma (or Fermi gas). As previously demonstrated~\cite{Bonatto2023}, under proper conditions, the wakefield driven by the laser within these nanostructured targets can be analytically estimated if an effective density is adopted. This approach enables the application of existing analytical expressions and knowledge developed for homogeneous plasmas to investigate the behavior of nanostructured materials.

An earlier numerical study~\cite{Bonoiu2023} demonstrated a gradient of approximately $4.8$ TeV/m in a multilayer graphene target, using an ultraviolet (100 nm wavelength), 3 fs-long laser pulse as the driver. While ultraviolet lasers are more available than X-ray lasers, they are not as widely adopted as infrared sources. In this work, with the aid of the effective density approach, we have identified suitable conditions for achieving laser wakefield acceleration (LWFA) using an infrared (800~nm) laser. This advancement leverages existing laser technology, making it more feasible to generate TV/m wakefields without the need of attosecond X-ray or ultraviolet lasers, thereby broadening the potential for experimental investigations and applications. Recent experimental results have demonstrated electron acceleration — with energies up to 1.2 MeV — using an 800 nm laser in a nanostructured target\cite{Dulat2024}. While this represents a significant advance in the field, using the effective density approach allows target geometries to be tuned for the formation of wakefield bubbles driven by an 800 nm laser pulse. These bubbles enable electron acceleration via self-injection, similar to processes observed in gaseous plasmas—albeit with TeV/m accelerating gradients due to the high effective density achieved by these solid nanostructured targets—and produce electron bunches with higher energy and charge over propagation distances of only a few micrometers. To our knowledge, this is the first work to report numerical results of TeV/m-level accelerating gradients driven by an infrared (800~nm), $2.67$~fs-long (FWHM) laser pulse in a solid nanostructured target.

\begin{figure*}[h]
\centering
\includegraphics[width=0.6\textwidth]{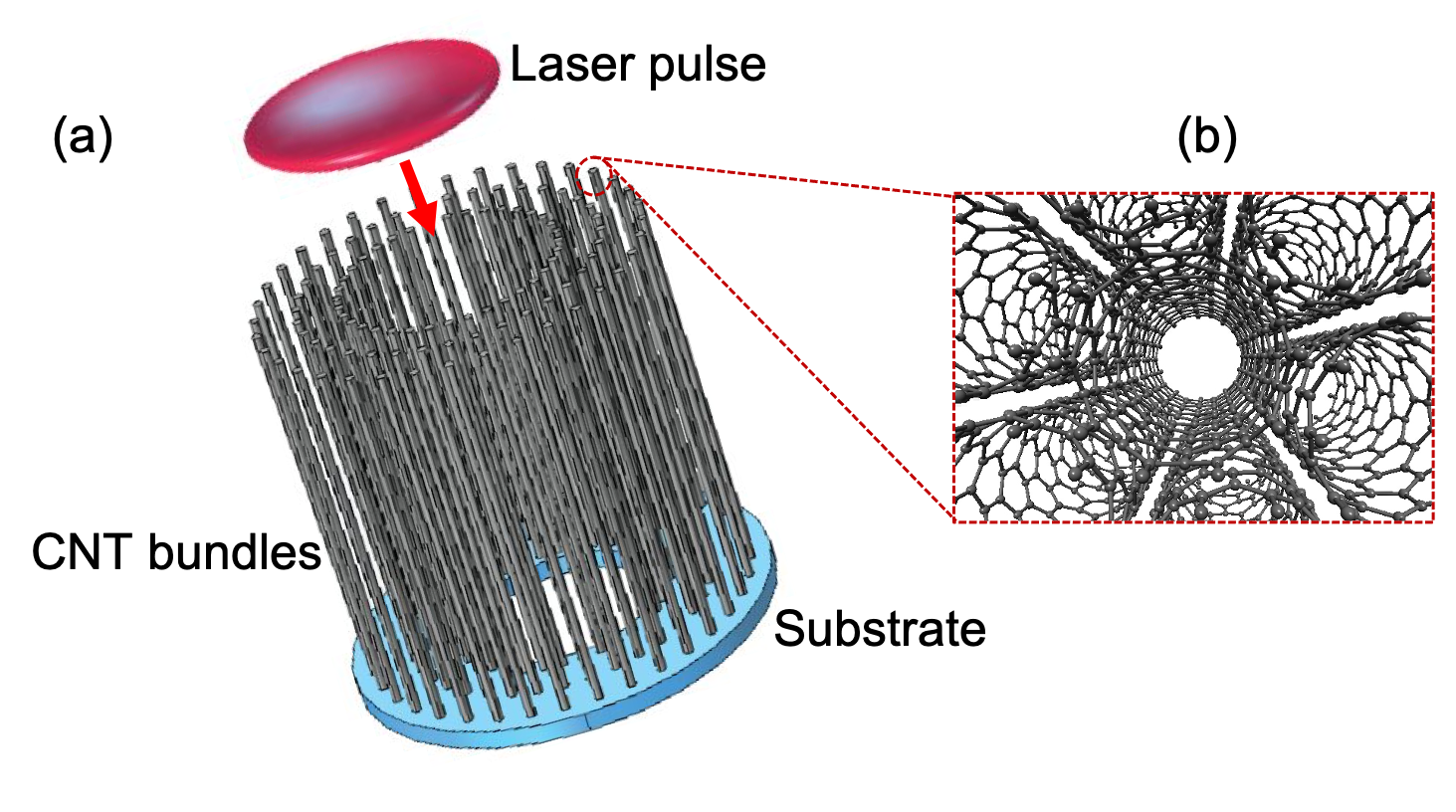}
\caption{(a) Schematic of a target based on CNT bundles vertically grown on a substrate with a circular aperture to allow the transmission of laser pulses throught it. (b) Zoomed-in view of the cross-sectional perspective of a CNT bundle.}
\label{fig:scheme} 
\end{figure*}

While LWFA research is currently dominated by meticulously tailored gaseous targets \cite{Gonsalves2019}, solid-state plasmas may soon become an alternative, due to their inherent advantages such as higher electron density and wider topological flexibility. It is possible for example to prepare hollow targets with controllable effective plasma density. Carbon nanomaterials such as graphene \cite{Bonoiu2023} and CNTs are good candidates due to the recent progress in their manufacturing techniques. This work considers 25 nm-thick bundles (ropes) of CNTs \cite{Thess1996} rather than large volumes (forests) of densely packed CNTs. Considering that a CNT bundle may contain tens or hundreds of tubes and inherent voids, it is reasonable to assume that the density of atoms is in the order of $10^{22} \, \text{cm}^{-3}$. A target can be manufactured distributing CNT bundles in concentric shells, as shown in Figure \ref{fig:scheme}, with an effective plasma density of $10^{20} \, \text{cm}^{-3}$.

In this article, our study focuses on resonant laser wakefield acceleration (LWFA) in solid-state targets composed of CNT bundles. It is important to distinguish this mechanism from the so-called plasmonics or plasmon acceleration scheme \cite{Fedeli2016, Sgattoni2016, Aakash2021, Aakash2023}. While plasmon acceleration primarily involves the coherent oscillation of surface electrons, leaving most of the target intact, resonant wakefield acceleration relies on the complete ionization and displacement of a substantial number of electrons throughout the entire target.

\section{Methods}
In this study, fully 3D particle-in-cell (PIC) simulations were performed using the PIConGPU code \cite{Burau2010}. The simulations were conducted on a machine with an Intel Core i9 processor with 16 cores, 64 GB of RAM, equipped with two Nvidia RTX 5500 GPUs, each with 16 GB of memory. The simulation domain consisted of $3 \times 10^8$ mesh cells, each measuring 25 \si{\nm}, with one macroparticle per cell. An 800 \si{\nano\m} wavelength laser pulse was propagated along the $y$-axis, corresponding to the longitudinal axis of the targets. The pulse has been assumed to have a Gaussian profile in both longitudinal and transverse directions, with circular polarization. The transverse electric field components were represented as horizontal ($E_z$) and vertical ($E_x$) fields. 

The primary objective of this study is to demonstrate that sufficiently intense infrared laser pulses can inject and accelerate electrons within ordered CNT structures. Bulkier targets typically cause rapid depletion and absorption of infrared laser pulses. To mitigate this issue, the targets have been designed as hollow tubes with walls composed of CNT bundles.

\begin{figure*}[!t]
\centering
\includegraphics[width=0.8\textwidth]{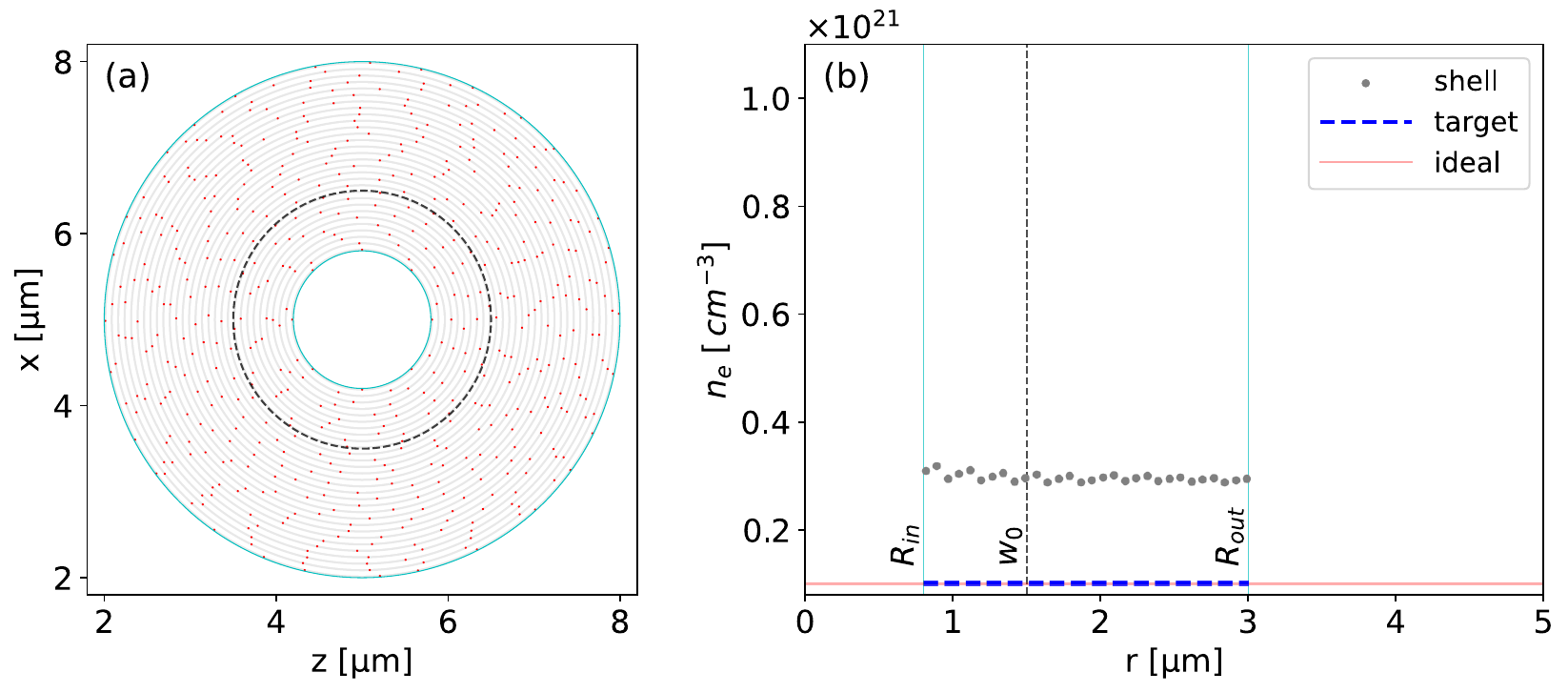}
\includegraphics[width=0.8\textwidth]{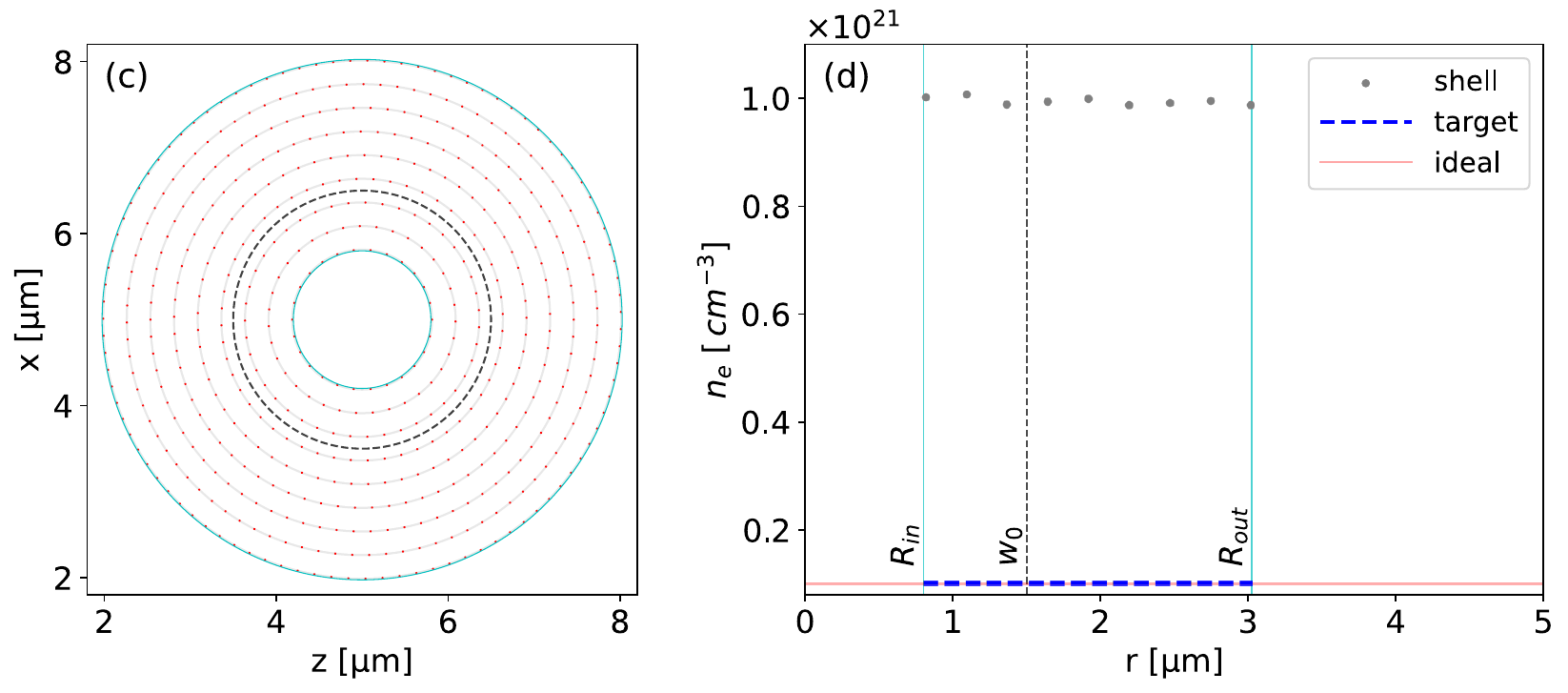}
\caption{(a) Cross-sectional view of CNT targets and the corresponding plasma density: (a-b) Transverse view of a target made of 535 CNT bundles randomly distributed in 30 shells with gap of 50 nm between shells, and the corresponding plasma density for each shell and (gray dots) for the whole target (dashed blue line); (c-d) Transverse view of a target made of 546 bundles neatly aligned in 9 shells with a gap of 250 nm between shells, and the corresponding plasma density for each shell (gray dots) and for the whole target (dashed blue line). The black dashed circle indicates the laser spot size ($w_0$), and both the inner ($R_{in}$) and the outer ($R_{out}$) radius of the target are indicated with a light blue line.}
\label{fig:target} 
\end{figure*}

Figure~\ref{fig:target} presents transverse and radial cross-sections of two distinct target configurations, both designed to achieve an effective density of $10^{20}\,\si{\per\cm\cubed}$, indicated by red lines (labeled as ``ideal'') in panels (b) and (d). In Fig.~\ref{fig:target}(a), the CNT bundles, represented as red dots, are randomly distributed across concentric shells denoted by grey lines, whereas in Fig.~\ref{fig:target}(c) these bundles are evenly distributed along the shells. The effective density for each shell, determined by the number of bundles, is depicted as grey dots in Fig.~\ref{fig:target}(b) and (d). In both configurations, the bundle distribution within the shells was chosen to ensure that the overall target's effective density (represented by the dashed blue lines) aligns with the desired (ideal) value.
%
%
%

In modern laboratories, the fabrication of these types of CNT structures is feasible using various methods, such as chemical vapor deposition (CVD), arc discharge, or template-assisted growth. Each method provides a unique level of control over the properties of the resulting CNTs and their bundles \cite{See2007, Sarasini2022,Schifano2023, Yadav2024}. Ultimately, current nanofabrication techniques offer the precision and flexibility needed to achieve the desired effective density parameters and target geometry.

As mentioned earlier, an effective plasma density of $10^{20}\,\si{\per\cm\cubed}$ is used in the PIConGPU simulations of this study. This density is well-suited to the laser's wavelength, as the plasma-to-laser frequency ratio remains in the underdense regime even when the six valence electrons of the carbon atoms are ionized. For comparison, Table~\ref{tab:plasma-laser-params} provides the plasma frequencies and wavelengths associated with three distinct effective densities — $10^{20}$, $10^{21}$, and $10^{22}\,\si{\per\cm\cubed}$ — as well as their ratios to the corresponding laser quantities. For the laser pulse to propagate through the plasma, the underdense condition ($\omega_p/\omega_0 < 1$) must be satisfied. As shown in Table~\ref{tab:plasma-laser-params}, while this condition is met for effective densities of $10^{20}$ and $10^{21}\,\si{\per\cm\cubed}$, for $10^{22}\,\si{\per\cm\cubed}$ the plasma becomes overdense even if a single ionization is assumed for the carbon ions. As the laser interacts with the CNT target, the carbon atoms undergo multiple ionization states. During the initial tens of femtoseconds, ionization advances rapidly, resulting in a plasma density increase in the interaction core that reaches $6\,n_e$. This escalation raises the plasma-to-laser frequency and wavelength ratios, $\omega_p/\omega_0$ and $\lambda/\lambda_p$, respectively, by a factor of $\sqrt{6}$. Following this scaling, the only configuration among those presented in Table~\ref{tab:plasma-laser-params} that satisfies the underdense condition required for the laser driver used in this work has an effective density of $10^{20}\,\si{\per\cm\cubed}$. Therefore, this plasma density was selected to ensure proper laser pulse propagation under any carbon ionization scenario.


\begin{table}[!t]
    \centering
    \caption{Comparison of plasma parameters considering a 800 \si{\nano\meter} laser pulse with $\omega_0=2.355\,\si{\radian - \peta\hertz}$ for the case of three different effective plasma densities, considering a single-level ionization of the carbon atoms.}
    \begin{tabular}{lcccl}
        \hline
        Parameter & \multicolumn{3}{c}{Plasma properties} & Unit\\ \hline
        Effective plasma density, $n_e$ & $\rule{0pt}{2.5ex}10^{20}$ & $10^{21}$ & $10^{22}$ & $\si{\per\cm\cubed}$ \\ 
        Plasma angular frequency, $\omega_p $& 0.564 &  1.784 &  5.641  &\si{\radian - \peta\hertz}  \\ 
        Plasma-to-laser frequency ratio, $\omega_p / \omega_0$ & 0.240 & 0.758 & 2.396 & -  \\ 
        Plasma wavelength, $\lambda_p$ & 3.339 & 1.056 & 0.334 &  \si{\um} \\ 
        Laser-to-plasma squared wavelength ratio, $\left(\lambda/\lambda_p\right)^2$ & $5.741 \times 10^{-2}$ & $5.741 \times 10^{-1}$ & 5.741 & - \\ \hline
    \end{tabular}
    \label{tab:plasma-laser-params}
\end{table}

In the proposed target configurations, the bore is an important feature, playing a role similar to the short-lived plasma channels \cite{Picksley2020} generated in gaseous targets. In addition to guiding the laser pulse over several Rayleigh lengths, it also provides the ion lattice along which the electron bunch is accelerated. In this study, simulations were performed using laser pulses with the parameters listed in Table~\ref{tab:laser-params}. A bore radius ($R_{in}$) of approximately half the laser spot size ($w_0$) was adopted, ensuring a ratio of $R_{in}/w_0 \approx 0.5$. While the choice of $w_0 = 1.5 \, \si{\micro m}$ was driven by computational limitations, the self-injection and acceleration scheme described in this work is expected to remain effective for larger spot sizes, provided this ratio is preserved. Similarly, while the full pulse length $ \Delta t $ represents 3 laser cycles, slightly longer pulses can be utilized. However, only peak intensity values on the order of $10^{21} \, \si{\W/\cm\squared}$ lead to significant target ionization and consequently electron self-injection and resonant acceleration. 
%

\begin{table}[!t]
    \centering
    \caption{Laser parameters.}
    \begin{tabular}{lccl}
        \hline
        {Parameter}  & \multicolumn{2}{c}{Value} & Unit  \\ \hline
        Wavelength, $\lambda$ & \multicolumn{2}{c}{800} & \si{\nm} \\
        Period, $T$  & \multicolumn{2}{c}{2.67} & \si{fs} \\ 
        FWHM pulse duration, $\Delta t$ & \multicolumn{2}{c}{2.67} & \si{fs} \\
        Spot size, $w_0$ & \multicolumn{2}{c}{1.5} & \si{\um} \\
        Peak intensity, $I_0$ & $10^{20}$ & $10^{21}$ &  \si{\W/\cm\squared} \\
        Potential vector, $a_0$ & 6.8 & 21.6 & - \\
        Pulse energy, $E$ & 10 & 100 & \si{\mJ} \\
    \hline
   \end{tabular}
    \label{tab:laser-params}
\end{table}

The laser and target parameters have been optimized to enhance the wakefield acceleration and maximize both the kinetic energy and charge of the extracted accelerated bunch. For instance, optimal values have been obtained for the ratio $R_{in}/w_0 \approx 0.53$. As aforementioned, the CNT bundles do not need to be arranged in strictly equidistant order, provided they achieve the required effective plasma density. This characteristic could offer advantages in terms of fabrication. In the next section, to analyze the simulation results, macroparticles were selected through backtracking, enabling the reconstruction of their dynamics during the interaction. In the next sections only results for the target configuration of Fig.~\ref{fig:target} (a, b) are shown. 


\section{Results}

\subsection{Self-injection and acceleration}
\begin{figure*}[!t]
\centering
\includegraphics[width=0.8\textwidth]{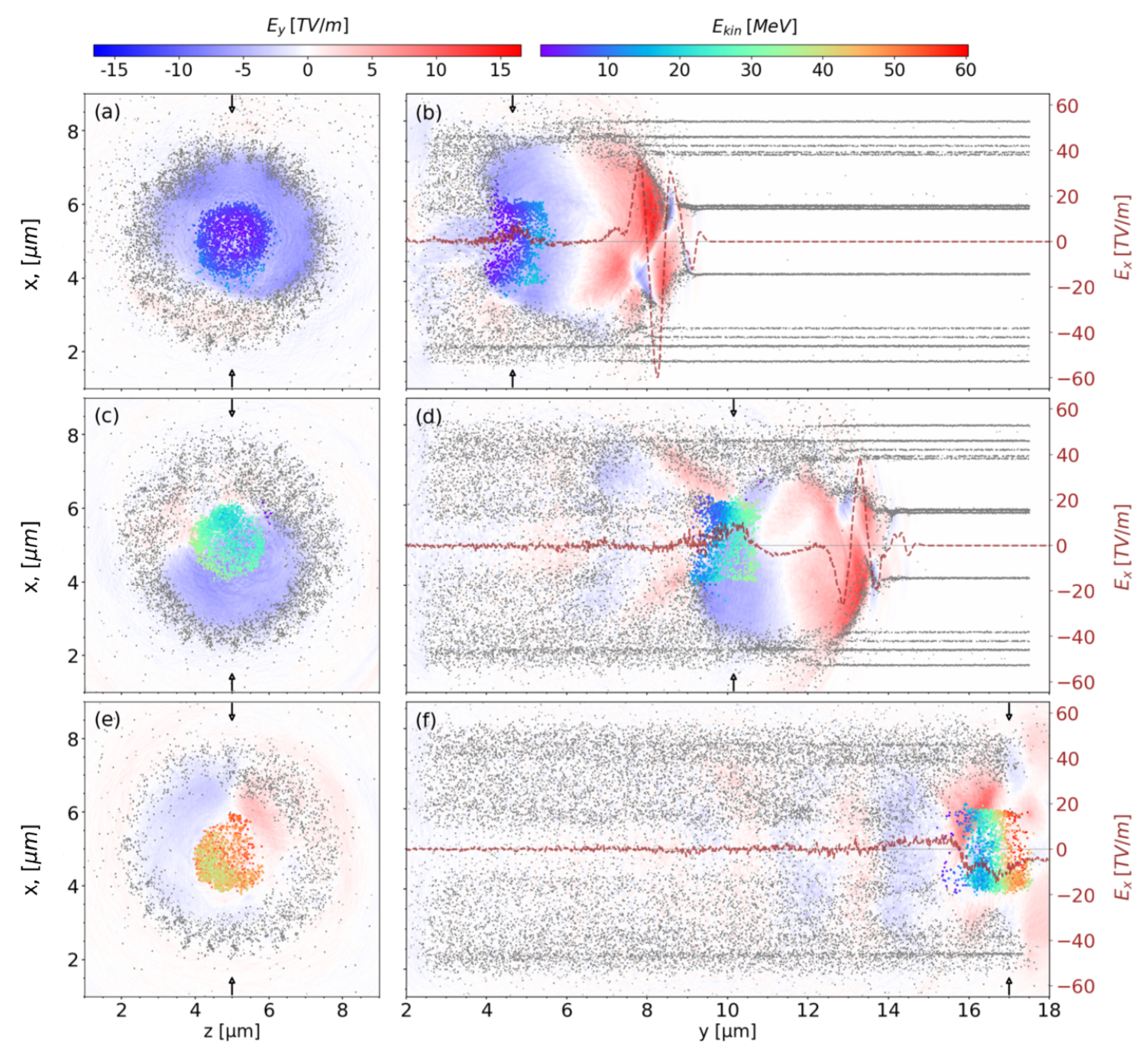}
\caption{Electron macroparticles shown as gray dots and the longitudinal electric field shown as a colour density plot for the target with constant effective density $n_e\simeq 10^{20}$ cm$^{-3}$, considering a laser pulse length $\Delta t=8$ fs (3 cycles) and intensity $I_0=10^{21}$ W/cm$^2$: (a-b) $t/T = 11$; (c-d) $t/T = 18$; (e-f) $t/T = 25$.}
\label{fig:electrons} 
\end{figure*}

Figure \ref{fig:electrons} shows the simulation results using the laser parameters listed in Table \ref{tab:laser-params} and the target depicted in Figure \ref{fig:target} (a-b). Upon the laser pulse striking the target, it ionizes the interaction region, causing electrons to be repelled toward the outer shells. This process generates a moving wakefield bubble, akin to those observed in LWFA gas-jet plasmas. Electrons are then self-injected at the rear of the bubble, where they experience longitudinal electric fields in the TV/m range.

Figures \ref{fig:electrons} (a-b) show a simulation snapshot of the process when the laser is nearly centered on the target. These figures depict a 5-\si{\um}-wide wakefield bubble, with its electric field represented on a blue-white-red color scale. The bubble contains an electron bunch, whose particles' kinetic energy is indicated by a rainbow color scale, applied only to the backtracked macroparticles that ultimately form the accelerated bunch at the end of the simulation. These particles are self-injected and accelerated at the bubble's tail, in the $E_y < 0$ region. The gray dots in the figures represent other system macroparticles that did not meet the selection criteria. Figure \ref{fig:electrons} (b) shows that the macroparticles ahead of the laser pulse remain organized, as the gray dots align in a carbon nanotube-like structure. However, once the laser pulse passes, these macroparticles interact with it, forming a disordered plasma structure in its wake. This disorganization, particularly visible in the region $2\,\si{\um} \leq y \leq 4\,\si{\um}$, supports the formation of a well-defined wakefield bubble.

Figures \ref{fig:electrons} (c-d) show the moment when the wakefield bubble passes the center of the target. At this point, the bubble contracts, and, as seen in Figure \ref{fig:electrons} (c), the bunch and bubble slightly diverge, moving out of phase toward the $E_y = 0$ line. The macroparticles within the bubble continue to gain energy, but the head of the bunch gains more than the tail, reaching 30 \si{\MeV} shortly after passing the midpoint of the target. In Figures \ref{fig:electrons} (e, f), the wakefield bubble reaches the right edge of the target, and the bunch begins to lose energy as it exits. By the end of the simulation, the bunch contains electrons with energies ranging from 0.04 \si{\MeV} to a maximum of 63.6 \si{\MeV}, with an average energy of 27.9 \si{\MeV} and a total charge of 867.5 \si{\pico\coulomb}. Considering the total distance traveled is just 15 \si{\um}, the average and maximum acceleration gradients are 1.86 \si{\TeV/\m} and 4.24 \si{\TeV/\m}, respectively. 

Figures \ref{fig:electrons} (a, c, e) present transverse cuts at the positions marked by arrows in panels (b), (d), and (f). These cuts show that, in cross-section, the particles have similar energies, and the electric field rotates, causing the slight divergence of the particles. In panel (c), a more densely populated area of particles is visible near $x=6 \,\si{\um},\,z=5\,\si{\um}$. A similar effect is seen in panel (e) at $x=4 \,\si{\um},\,z=4.5\,\si{\um}$, indicating that the bunch also rotates around the propagation axis as it is accelerated.

\begin{figure*}[!t]
\centering
\includegraphics[width=0.8\textwidth]{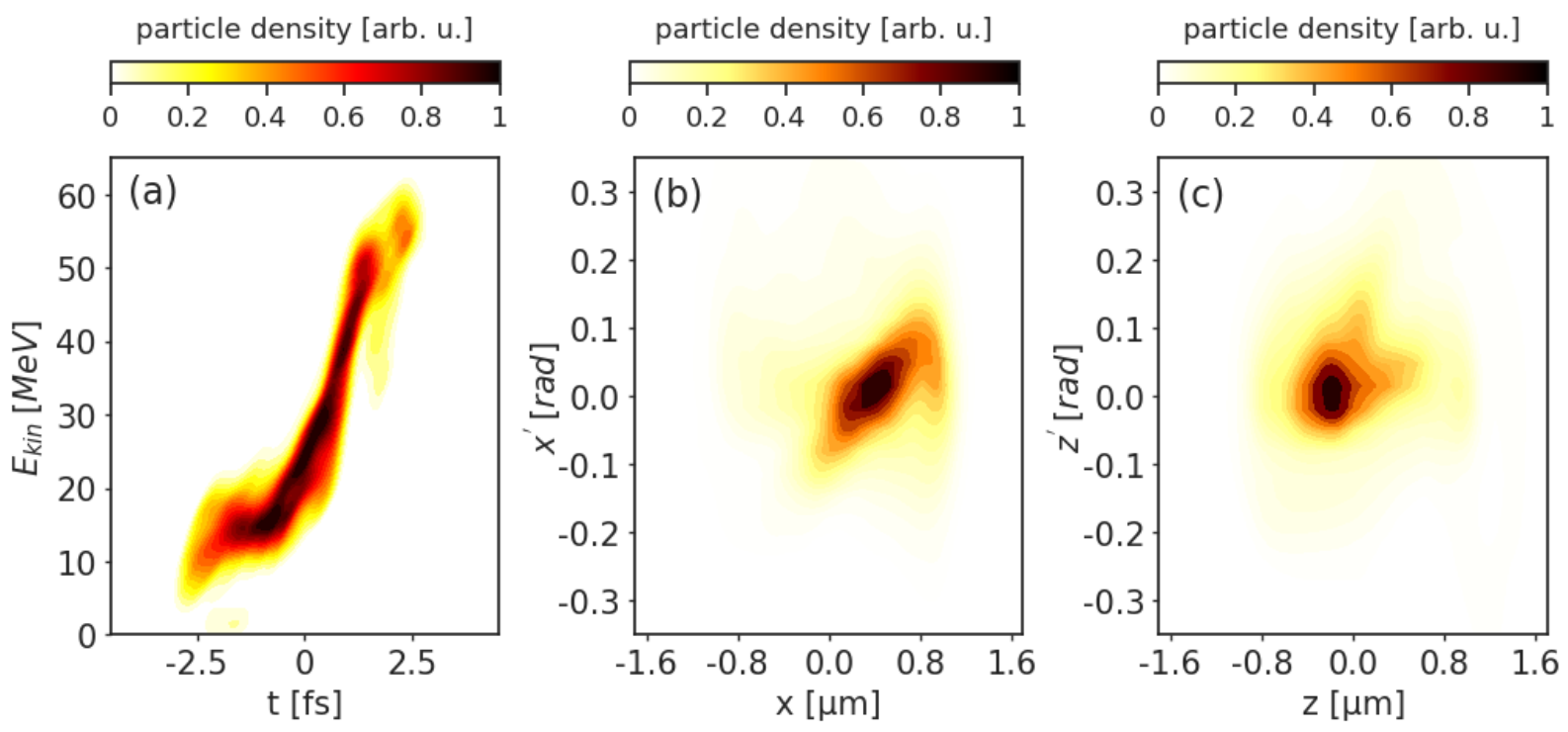}
\caption{(a)
longitudinal, (b) vertical, and (c) horizontal phase spaces at extraction, for a constant effective density of $n_e \simeq 10^{20} \, \si{\per\cm\cubed}$, and a laser pulse with $\Delta t \, \mathrm{(FWHM)} = 2.67\,\si{\femto\s}$, $I_0=10^{21} \, \si{\W/\cm\squared}$ and $R_{in}/w_0 = 0.53$ at $t/T = 26$.}
\label{fig:trace_space} 
\end{figure*}

Figure \ref{fig:trace_space} (a-c) presents colour density plots of the 6D phase space, showing the longitudinal projection in panel (a) and the vertical and horizontal projections in panels (b) and (c), respectively. The longitudinal phase space reveals a large energy spread, with particles ranging from 0.04 to 63.6 \si{\MeV}, and shows that the selected bunch fits within a 5 \si{\femto\second} time frame. The transverse phase spaces indicate significant divergence, with an amplitude of approximately 0.3 \si{\radian}, and show that the bunch is off-center relative to the target axis. This also suggests that the particles undergo transverse oscillations, possibly spiraling around the axis. Table \ref{tab:exit-final} presents a full characterization of the accelerated beam at the extraction point, for two different laser intensities. It is worth noting that gradients on the order of TV/m are generated, enabling the acceleration of self-injected electrons to kinetic energies of tens of MeV over distances of just 15 \si{\um}. The charge obtained is also remarkable, exceeding 100 pC, concentrated in short bunches of FWHM length on the order of 1 fs. However, these bunches are extracted with relatively high energy spread and transverse divergence not smaller than 130 mrad. Possible solutions to mitigate the large divergence could involve introducing a radial gradient in the plasma density, as this parameter directly influences the laser's phase velocity, which the electron bunch must match for optimal acceleration. 

\begin{table}[!htb]
    \centering
    	\caption{Bunch parameters at extraction for laser intensities $I_0$ = \SI[per-mode=symbol]{d20}{\watt\per\square\cm} and $I_0$ = \SI[per-mode=symbol]{d21}{\watt\per\square\cm}.}

    \begin{tabular}{lccl}
        \hline
        Parameter  & \multicolumn{2}{c}{Value at} & Unit  \\ 
        {} & $I_0=10^{20}$ &  $I_0=10^{21}$ & W/cm$^2$ \\ \hline
        Charge, $Q$ & 0.12 & 0.87 & \si{\nC} \\
        Average kinetic energy, $E_{kin}$  & 10.63 & 27.91 & \si{\MeV} \\ 
        Average acceleration gradient, $E_{kin}/L$ & 0.71 & 1.86 & \si{\tera\electronvolt/\meter} \\
         FWHM bunch length, $\Delta t_b$ & 1.72 & 3.98 & \si{\fs} \\
         FWHM energy spread, $\Delta E$ & 115 & 105 & \% \\
         Normalized RMS longitudinal emittance, $\bar{\rule{0pt}{2.5mm}\varepsilon}_{||}$ & 0.06 & 0.52 & \si{\fs - \MeV} \\
         FWHM vertical size, $\Delta x$ & 1.43 & 1.17 & \si{\um}  \\
         FWHM vertical divergence, $\Delta x'$ & 0.22 & 0.13 & \si{\radian} \\
         Normalized RMS vertical emittance, $\bar{\rule{0pt}{2.5mm}\varepsilon}_{x}$ & 1.16 & 2.16 & $\pi\,$\si{\mm -\m\radian}   \\
         FWHM horizontal size, $\Delta z$ & 1.24 & 0.82 & \si{\um} \\
         FWHM horizontal divergence, $\Delta z'$ & 0.19 & 0.13 & \si{\radian}  \\
         Normalized RMS horizontal emittance, $\bar{\rule{0pt}{2.5mm}\varepsilon}_{z}$ & 1.30 & 2.19 & $\pi\,$\si{\mm -\m\radian} \\
    \hline
   \end{tabular}
    \label{tab:exit-final} 
\end{table}

\begin{figure*}[!t]
\centering
\includegraphics[width=0.8\textwidth]{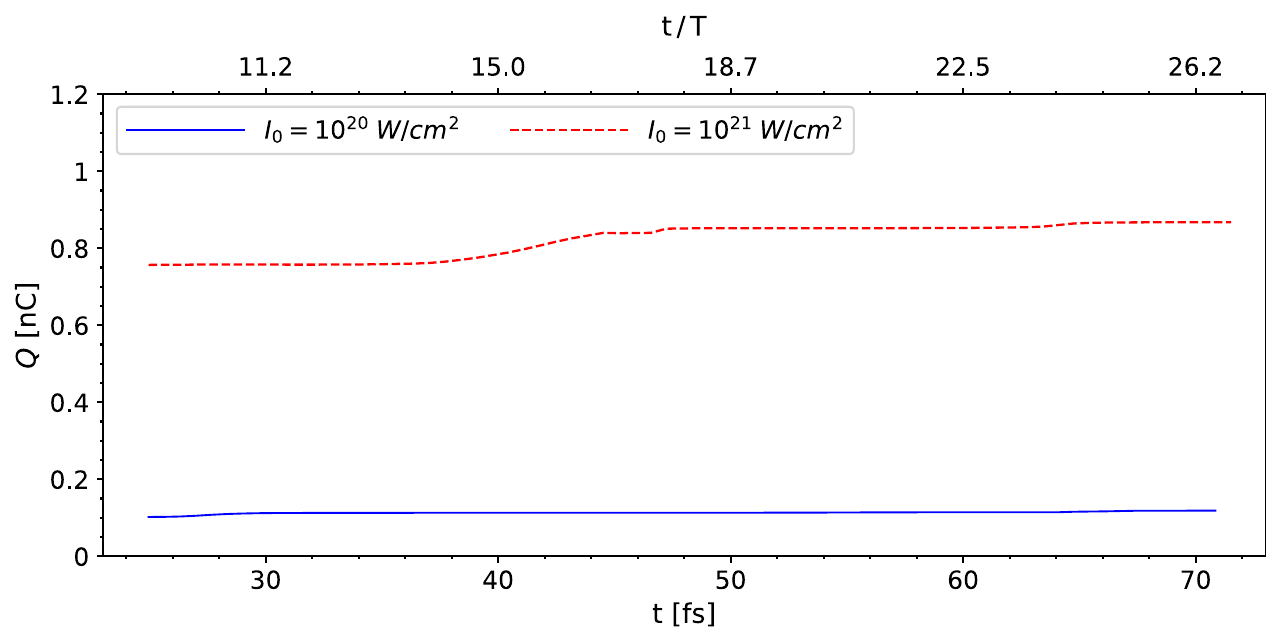}
\caption{Time evolution of the bunch charge $Q$ for two peak laser intensities, $I_0=10^{20}$ W/cm$^2$ and $I_0=10^{21}$ W/cm$^2$, shown in femtoseconds and normalized by the laser period $T$.}
\label{fig:combined-charge} 
\end{figure*}

There are significant differences between the bunch charge values at \SI[per-mode=symbol]{d20}{\watt\per\square\cm} and at \SI[per-mode=symbol]{d21}{\watt\per\square\cm}. As shown in Fig.~\ref{fig:combined-charge}, increasing the peak intensity by one order of magnitude results in a seven-fold increase in bunch charge. This difference can be attributed to the stronger longitudinal electric field $E_y$, which continuously pulls more electrons into the second half of the wakefield bubble. Additionally, the initial plasma shock wave at the left edge of the target plays a crucial role, as it triggers electron self-injection. Moreover, when the laser peak intensity is sufficiently high, the bunch also collects electrons from the front wall of the wakefield. The first half of the bubble contains positive longitudinal electric field $E_y$, and transverse electric fields $E_x$ and $E_z$ which exhibit mirror symmetry across the bubble midpoint relative to the fields in the second half. As a result, electrons from the front bubble wall may, in certain situations, be accelerated backward toward the bubble center. They are then captured by the second half of the bubble, which accelerates them forward. Consequently, the bunch acquires additional electrons, leading to an increase in its charge.    

\begin{figure*}[!t]
\centering
\includegraphics[width=0.8\textwidth]{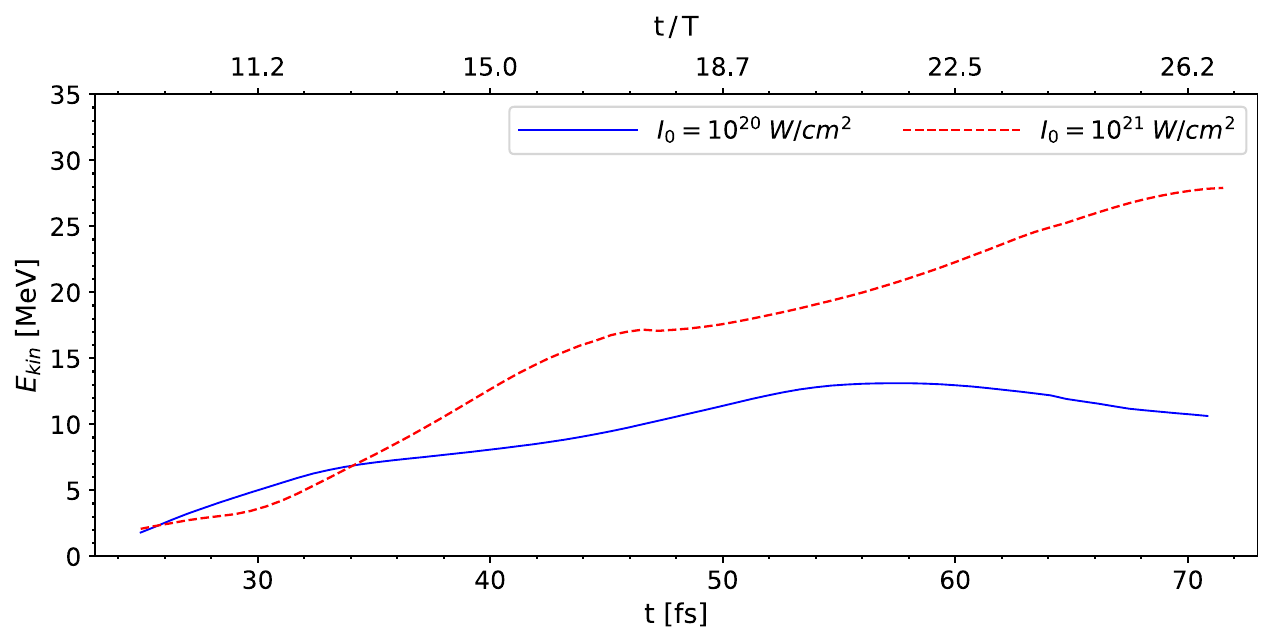}
\caption{Time evolution of the bunch kinetic energy $E_{kin}$ for two peak laser intensities, $I_0=10^{20}$ W/cm$^2$ and $I_0=10^{21}$ W/cm$^2$, shown in femtoseconds and normalized by the laser period $T$.}
\label{fig:combined-energy} 
\end{figure*}

Figure~\ref{fig:combined-energy} depicts the evolution of the average kinetic energy, $E_{kin}$, of the self-injected bunch during the laser-plasma interaction for both values of the laser peak intensity. As expected, for \SI[per-mode=symbol]{d20}{\watt\per\square\cm}, $E_{kin}$ grows slower and reaches a maximun before the bunch is extracted. This indicates that dephasing occurs, and some of the bunch electrons cross the wakefield bubble middle, at $t/T \simeq 20$, after which they experience decelerating electric fields ($E_y>0$). For \SI[per-mode=symbol]{d21}{\watt\per\square\cm} the energy gain rate is higher and dephasing is not evident. However, in this case, due to the intake of low-energy electrons, $E_{kin}$ shows a sudden drop at $t/T\simeq 17$. This is clearly correlated with sudden increase of charge observed in Fig.~\ref{fig:combined-charge}. 

\subsection{Induced magnetic field and focusing effects}
\begin{figure*}[!t]
\centering
	\includegraphics[width=0.8\textwidth]{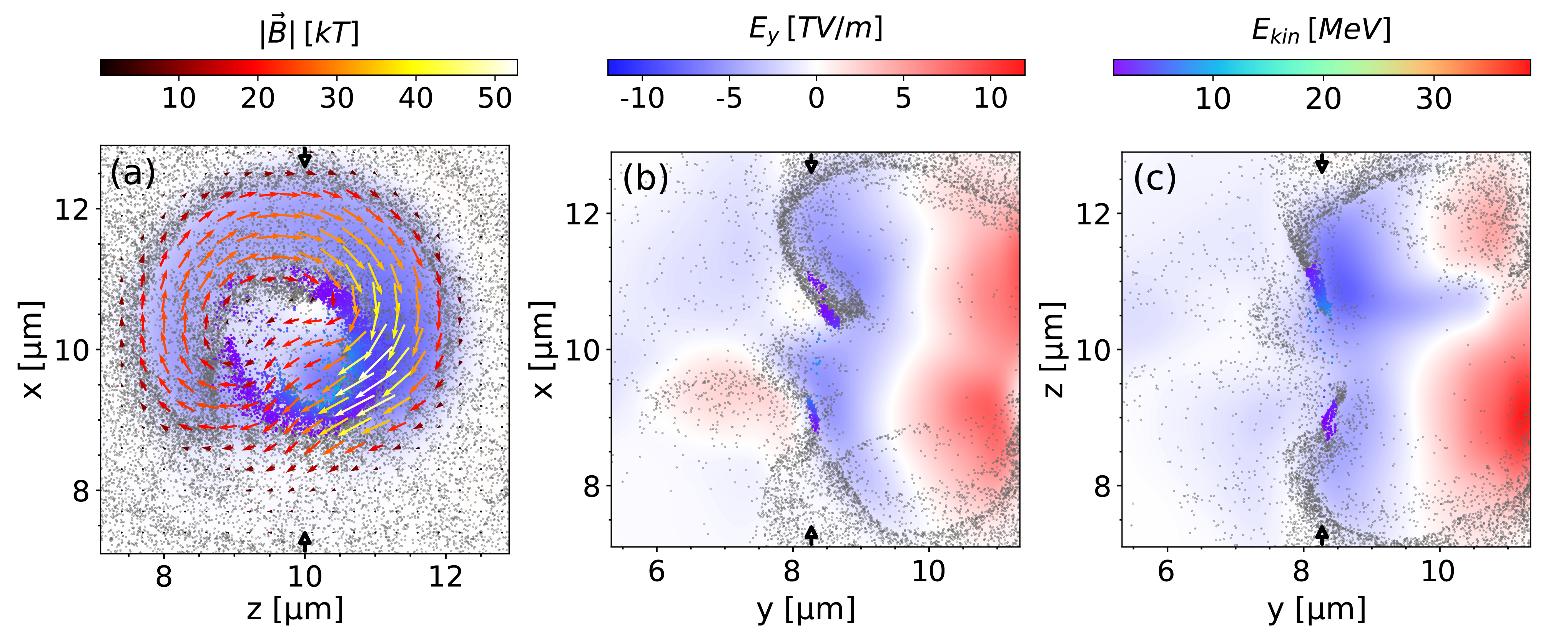}
	\caption{The electron bunch shown as an energy-coloured selection of macroparticles in the transverse and axial cut-planes along with the longitudinal electric field $E_y$ at $t/T=10$. Magnetic field vectors are shown in the $zx$-plane, with the macroparticles moving towards the reader. Thus, the longitudinal motion coupled with the azimuthal magnetic field has a focusing effect: (a) in the $zx$-plane (transverse); (b) in the $yx$-plane (vertical); and (c) in the $yz$-plane (horizontal).}
	\label{fig:magnetic_field}
\end{figure*}

At injection, the bunch experiences longitudinal accelerating electric fields whose intensity sometimes exceeds 5 TV/m, up to the moment when the wakefield bubble breaks. At the same time, the bunch is immersed in azimuthal magnetic field whose induction peaks at $\simeq 50$~kT. In consequence, this induced field has a focusing effect on the particles. This effect is shown in
Fig.~\ref{fig:magnetic_field}, where the electron dynamics and induced fields are monitored at $t/T=10$. 

It is worth mentioning that high-intensity magnetic fields are required in a number of applications, and the method described here could be considered a powerful source. Among others, an interesting application of the extremely high magnetic fields generated in laser-plasma interactions is their potential to wiggle electron beams and effectively produce an X-ray free-electron laser (FEL). For example, as estimated by Loeb and Eliezer\cite{Loeb1986}, a 150 MeV electron beam could generate coherent radiation with a 1 nm wavelength, provided it is subjected to a magnetic field as strong as 100 T. Due to their robustness and higher inertia compared to gaseous plasmas, carbon nanotube (CNT) targets may offer the advantage of sustaining high magnetic fields for extended durations.

\subsection{Effects on the ionic lattice}
\begin{figure*}[!t]
\centering
	\includegraphics[width=0.8\textwidth]{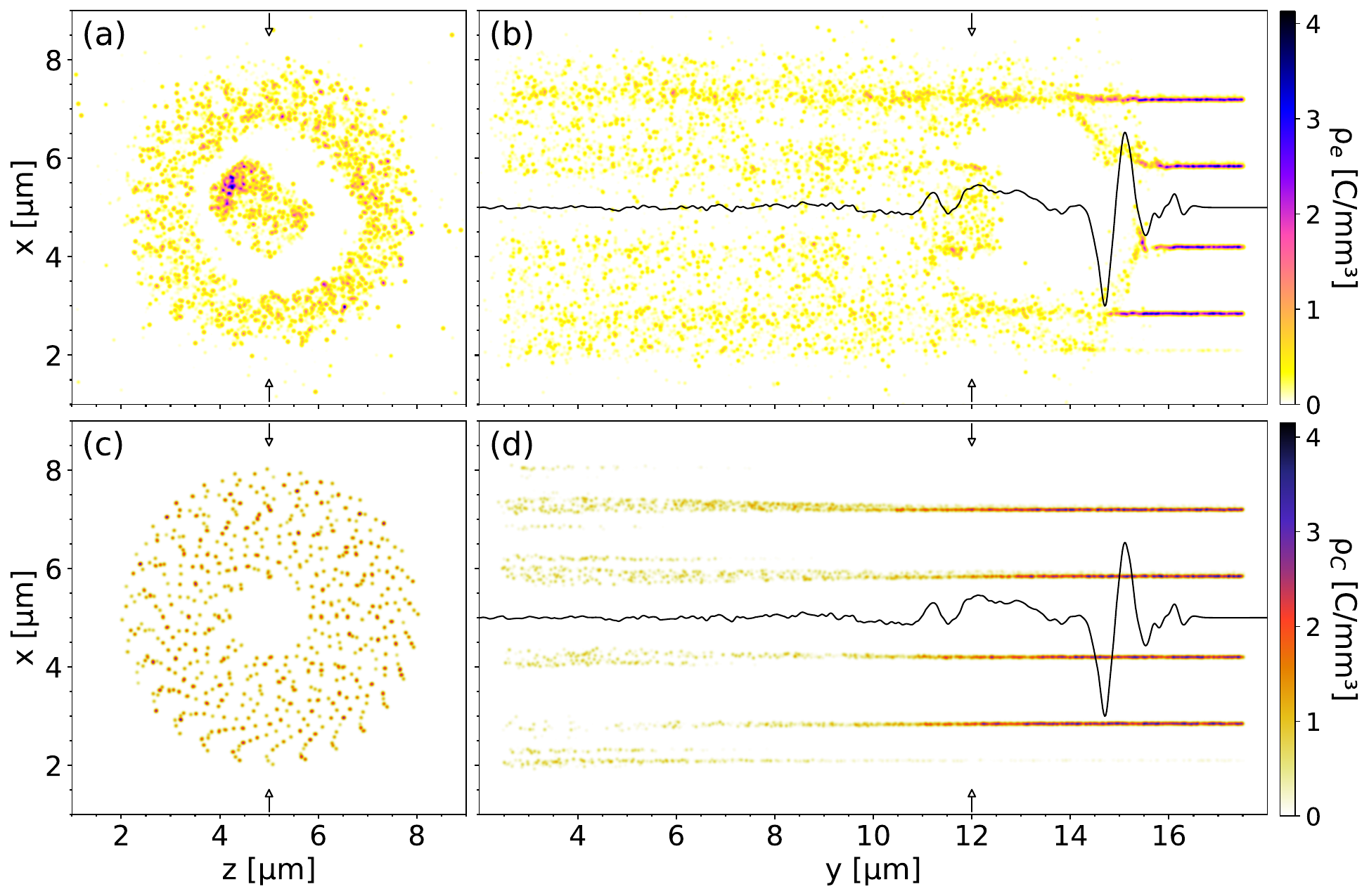}
	\caption{Synchronous comparison of the electron and carbon charge density in absolute values, with $I_0$ = \SI[per-mode=symbol]{d21}{\watt\per\square\cm}, $\Delta t$ = \SI{8}{\femto\second} (3 cycles), with $R_{in}$ = \SI{0.8}{\micro\metre}, $w_0$ = \SI{1.5}{\micro\meter} at $t/T$ = 20. The on-axis vertical electric field $E_x$, which comes mostly from the laser, is shown by the solid black line: (a-b) Transverse and axial view of the electron charge density $\rho_e$; (c-d) Transverse and axial view of the carbon charge density $\rho_C$.}
	\label{fig:charge-density-electrons-carbon}
\end{figure*}

Another aspect to be analyzed is the mobility of carbon ions under the action of Coulomb forces. Although $m_e$/$m_{C}\sim 10^{-4}$, given the high laser intensity and virtually instantaneous complete ionization within the laser pulse, an evaluation of the carbon ion displacement can be a reasonably safe validity assessment to justify the use of PIC codes in the simulation of this kind of solid-state structure. For this purpose, the charge density of both electrons ($\rho_e$) and carbon ions ($\rho_C$) was extracted from the laser-CNT interaction, utilizing target parameters similar to those adopted in the previous section. As the interaction progresses along the longitudinal ($y$) axis, the laser pulse completely ionizes the core region of the target and maintains a wakefield bubble. As shown in Fig.~\ref{fig:charge-density-electrons-carbon} (a-b), the bubble is void of electrons, and a bunch is self-injected from its back wall. It can also be seen that the laser pulse, shown by the solid black line, fits within the first half of the bubble. It is worth noting that at this point the laser pulse is depleted and, out of three cycles, only one is still clearly visible. It is important to assess how the carbon ion displacement affects the dynamics of the electron bunch. For this purpose, it is enough to plot the charge density of the carbon ions because, as with any PIC code, charge density is a derived magnitude computed from the macroparticle distribution. Thus, to the extent that the mesh and shape functions are acceptable, $\rho_C$ indicates the location of the carbon ions. As shown in Fig.~\ref{fig:charge-density-electrons-carbon} (c-d) there is negligible displacement of hundreds of \SI{}{\nano\metre} in the back of the bubble.

These results indicate that even if the carbon ions were considered part of a gas rather than a solid-state lattice, their displacement would be sufficiently small to validate the use of PIC simulations for this wakefield acceleration scheme. Moreover, it is evident that a short laser pulse of peak intensity $I_0$ = \SI[per-mode=symbol]{d21} {\watt\per\square\cm}, carrying \SI{100}{\milli\joule}, does not instantly destroy the carbon lattice. The inertia of the carbon ions allows for a $\sim$ \SI{20}{\femto\second} delay, enough for the electron bunch to be accelerated along a virtually unaltered ion lattice. This finding aligns with previous studies showing that wakefields in solid-state nanomaterials remain largely unaffected by ion dynamics \cite{hakimi2018}. Consequently, the PIConGPU code can be safely applied to simulate electron laser-driven acceleration in a solid-state lattice.

\section{Conclusions}

This work presents, for the first time, numerical results demonstrating that it is theoretically possible to achieve laser wakefield acceleration in a solid-state plasma with an 800 \si{\nano\m} (infrared) laser pulse. Using PIConGPU simulations, we demonstrate that  this laser can accelerate an electron pulse with a charge of 867 \si{\pico\coulomb} to an average energy of 27.9 \si{\MeV} within a 15 \si{\um}-long carbon nanotube bundle target, achieving acceleration gradients in the \si{\tera\electronvolt/\m} range. Furthermore, the obtained ultra-short pulse length ($\Delta t_b \lesssim 10$ fs) and relatively low transverse emittance ($\varepsilon_{x,z} \approx 2$ $\pi$ mm -- mrad), could make this acceleration technique suitable for interesting applications in the field of ultrafast science, e.g., for single-shot MeV ultrafast electron diffraction \cite{Salen2024}. It can also be an interesting mechanism to generate THz radiation for imaging applications \cite{Mittleman1999, Singh2013} with implications in various research areas such as: cell biology surface chemistry and condensed matter.

Our future research plan includes the further optimisation of the results presented in this article by the use of algorithms such as, for example, Bayesian Optimization \cite{Brochu2010}, which could be employed to further improve the beam energy, charge, and quality (divergence, energy spread), thereby enhancing its overall applicability.

Currently, the collaboration is preparing optimized targets for a proof-of-principle experiment, focusing on advanced manufacturing techniques. The required laser parameters, similar to those listed in Table~\ref{tab:laser-params}, are technically achievable in existing laser facilities, such as the Extreme Light Infrastructure (ELI)\cite{ELI} in Prague, Czech Republic, and the VEGA laser system at the Center of Pulsed Lasers (CLPU)\cite{CLPU} in Salamanca, Spain. If successful, this experiment could provide a novel alternative acceleration scheme and mark the first demonstration of resonant LWFA in a solid-state plasma. 

\black
\bibliography{sample}
\section*{Acknowledgements}
We acknowledge the support received from the PIConGPU software developers and the STFC CDT LIV.DAT under grant agreement ST/P006752/1. This work is supported by the Generalitat Valenciana under grant agreement CIDEGENT/2019/058, and by the the Fundação de Amparo à Pesquisa do Estado do Rio Grande do Sul (FAPERGS), grant 24/2551-0001552-3.

\end{document}